\documentclass{vldb}

\usepackage{datetime}
\usepackage{url}
\usepackage{multirow}
\usepackage{color}
\usepackage[pdfborder={0 0 0}]{hyperref}

\usepackage{array}
\newcolumntype{+}{>{\global\let\currentrowstyle\relax}}
\newcolumntype{^}{>{\currentrowstyle}}
\newcommand{\rowstyle}[1]{\gdef\currentrowstyle{#1}%
  #1\ignorespaces
}

\def\subclause{sub-clause}
\def\subclauses{sub-clauses}
\def\SUB{SUB}
\def\ms{\mathrm{ms}}
\def\ptile{\%\mathrm{ile}}

\makeatletter
\newcommand\footnoteref[1]{\protected@xdef\@thefnmark{\ref{#1}}\@footnotemark}
\makeatother

\begin{document}

%% IMPORTANT TODOs for final / next version:f
%% 
%% * When discussing the straightforward inverted index approach, clarify that this does not
%%   provide query suggestions, or only at a very high cost (ESTER does provide query suggestions
%%   faster than an inverted index, but still far from as fast as our new index).

\title{Broccoli: Semantic Full-Text Search at your Fingertips}
%\subtitle{Working Draft, Version \today, \currenttime}
%\subtitle{\color{red}\bf\Large Under review, please do not circulate!}

%% \title{A Sample {\ttlit ACM} SIG Proceedings Paper in LaTeX
%% Format\titlenote{(Does NOT produce the permission block, copyright information nor page numbering). For use with ACM\_PROC\_ARTICLE-SP.CLS. Supported by ACM.}}
%% \subtitle{[Extended Abstract]


%% Commented out for blind submission.
\numberofauthors{1}
\author{
\alignauthor
Hannah Bast, Florian B\"{a}urle, Bj\"{o}rn Buchhold, Elmar Haussmann\\[1mm]
%{(Note: names are in alphabetical order)}\\[1mm]
       \affaddr{Department of Computer Science}\\
       \affaddr{University of Freiburg}\\
       \affaddr{79110 Freiburg, Germany}\\
       \email{\{bast,baeurlef,buchholb,haussmann\}@informatik.uni-freiburg.de}
}
%\numberofauthors{1}
%\author{\alignauthor}

%% Remove the ACM copyright info.
%\toappear{}

\maketitle
\begin{abstract}
We present Broccoli, a fast and easy-to-use search engine for what we call
semantic full-text search.
Semantic full-text search combines the capabilities
of standard full-text search and ontology search.
The search operates on four kinds of objects: ordinary words (e.g., \emph{edible}), classes (e.g., \emph{plants}), instances (e.g., \emph{Broccoli}), and relations (e.g.,
\emph{occurs-with} or \emph{native-to}).
Queries are trees, where nodes are arbitrary bags of these objects, and arcs are relations.
The user interface guides the user in incrementally constructing such trees by instant
(search-as-you-type) suggestions of words, classes, instances, or relations that lead to good hits.
Both standard full-text search and pure ontology search are included as special cases.
In this paper, we describe the query language of Broccoli, the main idea behind a new kind of index
that enables fast processing of queries from that language as well as fast query suggestion, the natural language processing required, and the user interface.
We evaluated query times and result quality on the full version of the English Wikipedia (40 GB XML dump) combined with the YAGO ontology (26 million facts).
We have implemented a fully functional prototype based on our ideas
%\texttt{\href{http://broccoli.informatik.uni-freiburg.de}{broccoli.informatik.uni-freiburg.de}}.
and provide a web application to reproduce our quality experiments.
Both are accessible via \url{http://broccoli.informatik.uni-freiburg.de/repro-corr/} .
%\url{http://vldb2013-506.hopto.org} .
\end{abstract}

%\vspace{-1mm}
%
%% A category with the (minimum) three required fields
%\category{H.3.1}{Information Storage and Retrieval}{Context Analysis and Indexing}[Indexing
%methods, Linguistic processing]
%\category{H.3.3}{Information Storage and Retrieval}{Information Search and Retrieval}[Query formulation,
%Retrieval models, Search process]
%\category{H.5.2}{Information Interfaces and Presentation}{User Interfaces}[Theory and methods]
%%A category including the fourth, optional field follows...
%%% \category{D.2.8}{Software Engineering}{Metrics}[complexity measures, performance measures]
%
%\vspace{-1mm}
%
%\terms{Algorithms, Design, Experimentation, Human Factors, Performance}
%
%\vspace{-1mm}
%
%
%\keywords{Semantic full-text search, search as you type, Wikipedia}
%%% \keywords{ACM proceedings, \LaTeX, text tagging} % NOT required for Proceedings

%\vfill
\section{Introduction}\label{sec:intro}

In this paper, we describe a novel implementation of what we call \emph{semantic full-text search}.
Semantic full-text search combines traditional \emph{full-text search} with structured search in knowledge databases or \emph{ontology search} as we call it in this paper.

In traditional full-text search you type a (typically short) list of keywords and you get a list of
documents containing some or all of these keywords, hopefully ranked by some notion of relevance to your query.
For example, typing \emph{broccoli leaves edible} in a web search engine will return lots of web pages with evidence that broccoli leaves are indeed edible.

In ontology search, you are given a knowledge database which you can think of as a store of subject-predicate-object triples.
For example, \emph{Broccoli is-a plant} or \emph{Broccoli native-to Europe}.
These triples can be thought of to form a graph of entities (the nodes) and relations (the edges), and ontology search allows you to search for subgraphs matching a given pattern.
For example, find all plants that are native to Europe.

Many queries of a more ``semantic'' nature require the combination of both approaches. For example,
consider the query \emph{plants with edible leaves and native to Europe}, which will be our running example in this paper. A satisfactory answer for this query requires the combination of two kinds of information. First, a list of plants native to Europe. This is hard for full-text search but a showcase for ontology search, see above. Second, for each plant the information whether its leaves are edible or not. This kind of information can be easily found with a full-text search for each plant, see above. But it is quite unlikely (and unreasonable) to be contained in an ontology, for reasons explained in Section \ref{sec:related:ontoonly}.

\begin{figure*}[ht]
\includegraphics[width=\textwidth]{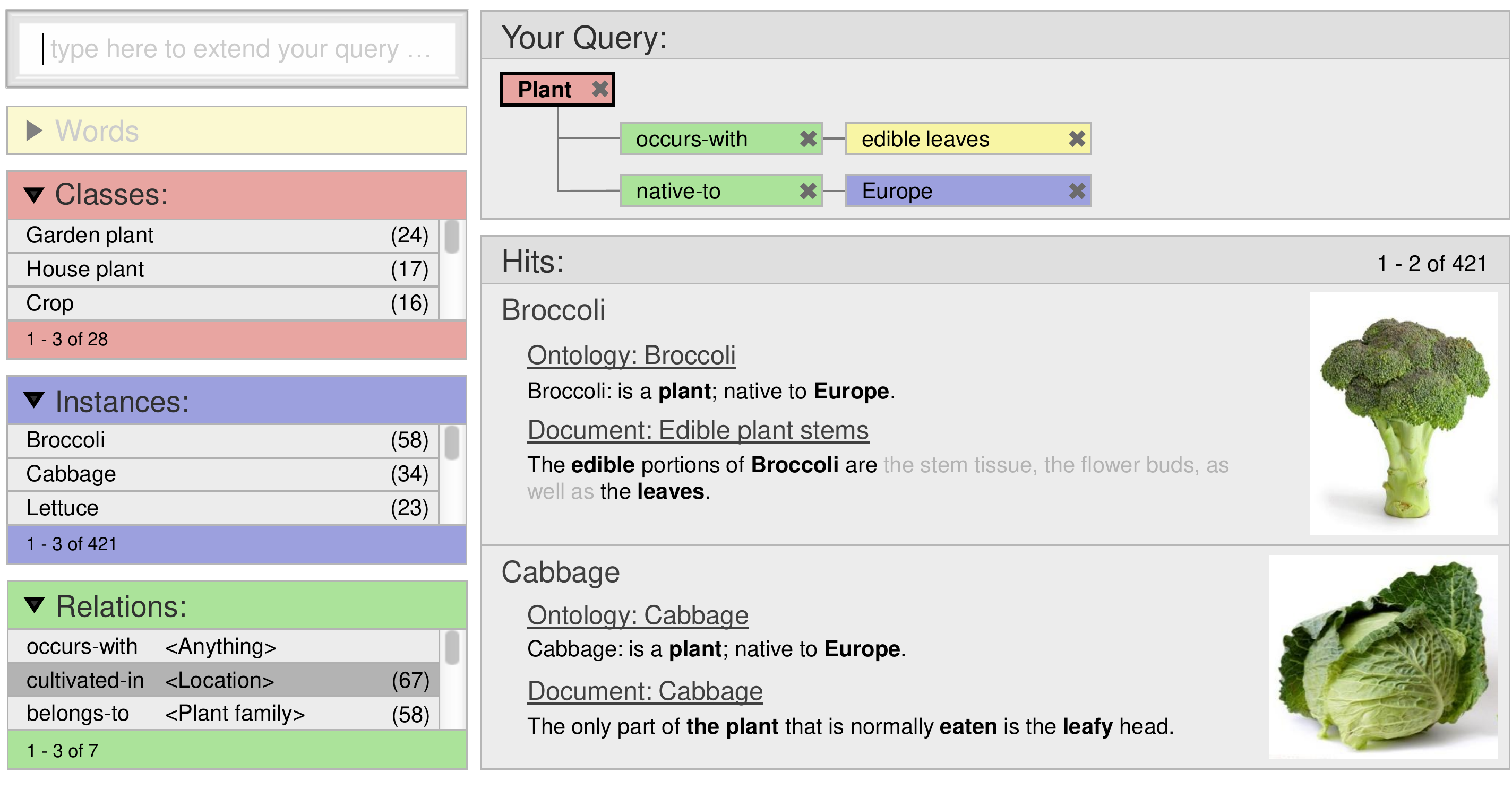}
\vspace{-7mm} %NOTE(Florian): reduce the space between the figure and its caption for the sake of space
\caption{A screenshot of the final result for our example query.
The box on the top right visualizes the current query as a tree.
There is always one node in focus (shown in bold), in this case, the root of the tree.
The large box below shows the hits grouped by instance (of the class from the root node) and ranked by relevance (if Broccoli is among the hits, we always rank it first).
Evidence both from the ontology and the full text is provided.
For the latter, a whole sentence is shown, with parts outside of the matching context grayed out.
With the search field on the top left, the query can be extended further.
%NOTE(Florian): shortened the rest of the description because of space
%The four boxes below provide context-sensitive suggestions that are updated after every keystroke.
%The suggestions depend on the current focus in the query, here:
%suggestions for sub\emph{classes} of plants,
%suggestions for \emph{instances} of plants that lead to a hit (those are exactly those from the hits box, and in the same order),
%suggestions for \emph{relations} to further refine the query.
%Word suggestions make no sense at this point in the query; see Figure \ref{figure:screenshot-2}.
%One of the suggestions is always highlighted, in this case the \emph{cultivated-in} relation.
%Pressing Return will extend the query by adding the highlighted suggestion.%
The four boxes below provide context-sensitive suggestions that depend on the current focus in the query, here:
suggestions for sub\emph{classes} of plants,
suggestions for \emph{instances} of plants that lead to a hit,
suggestions for \emph{relations} to further refine the query.
One of the suggestions is always highlighted, in this case the \emph{cultivated-in} relation.
It can be directly added to extend the query by pressing Return.%
\label{figure:screenshot-1}}
\end{figure*}

\begin{figure*}[ht]
\includegraphics[width=\textwidth]{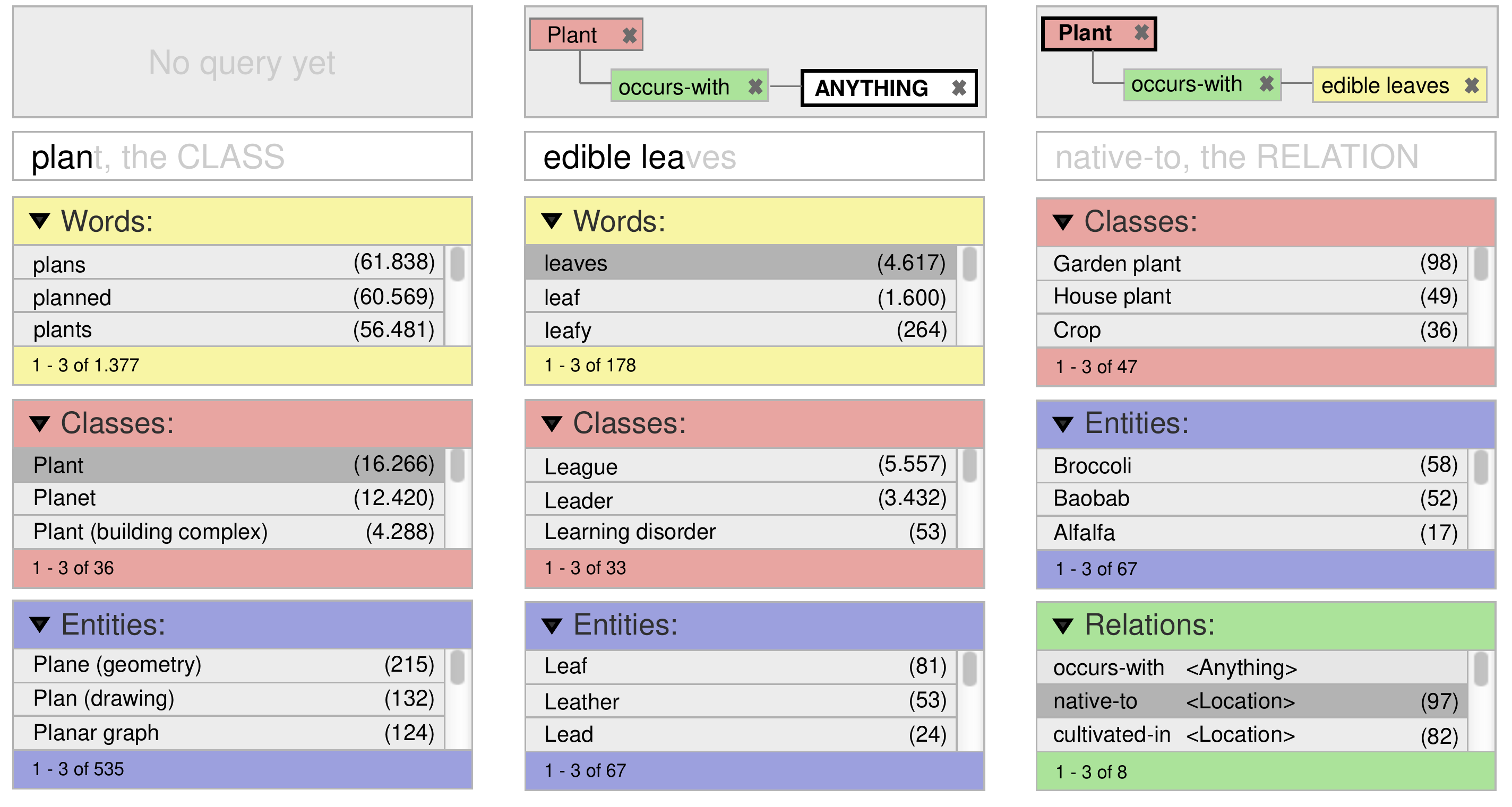}
\vspace{-6mm} %NOTE(Florian): reduce the space between the figure and its caption for the sake of space
\caption{Snapshots of the query, search field, and suggestion boxes for three stations in the construction of our example query.
Column 1: At the beginning of the query, after having typed \emph{plan}.
%NOTE(Florian): shortened the following two sentences because of space
%Column 2: After \emph{plant} has been selected (by pressing Return) and the \emph{occurs-with} relation has been added (by pressing Return again) and having typed \emph{edible lea}.
%Column 3: After having selected \emph{edible leaves} (by pressing Return).
Column 2: After the class \emph{plant} has been selected and the \emph{occurs-with} relation has been added and having typed \emph{edible lea}.
Column 3: After having selected \emph{edible leaves}.
%NOTE(Florian): shortened the following sentence because of space
%The focus automatically goes back to the root node, and the top relation that is not \emph{occurs-with} is pre-selected (we could select and add another \emph{occurs-with} if we wanted to though).%
The focus automatically goes back to the root node.%
\label{figure:screenshot-2}}
\end{figure*}

The basic principle of our combined search is to find \emph{contextual} co-occurrences of the words from the full-text part of the query with entities matching the ontology part of the query.
Consider the sentence: \emph{The stalks of \underline{rhubarb} are edible, but \underline{its} leaves are toxic}.
Assume for now that we can recognize entities from the ontology in the full text (we come back to this in Section \ref{sec:preproc:er}). In this case, the two underlined words both refer to \emph{rhubarb}, which our ontology knows is a plant that is native to Europe.
Obviously, this sentence should \emph{not} count as evidence that \emph{rhubarb leaves are edible}.
%We handle this by decomposing (after the entity recognition phase) each sentence into what we call its \emph{contexts}: the parts of the sentence that ``belong'' together.
We handle this by decomposing each sentence into what we call its \emph{contexts}: the parts of the sentence that ``belong'' together.
In this case \emph{the stalks of rhubarb are edible} and \emph{rhubarb leaves are toxic}.
An arc from the query tree now matches if and only if its elements co-occur in one and the same context.%; this will be explained in detail in Section \ref{sec:index:queryproc}.
%Then we search for co-occurrences of the elements from the query in each context separately.

Figures \ref{figure:screenshot-1} and \ref{figure:screenshot-2} show screenshots of our search engine in action for our example query.
The figures and their captions also explain how the query can be constructed incrementally in an easy way and without requiring knowledge of a particular query language on the part of the user.
%We encourage the reader to try our online demo on \url{http://broccoli.informatik.uni-freiburg.de}.
We encourage the reader to try our online demo that is accessible via \url{http://broccoli.informatik.uni-freiburg.de/repro-corr/} .
%\url{http://vldb2013-506.hopto.org} .

\newpage

\subsection{Our contribution}

%In this paper, we present Broccoli, a system for semantic full-text search that is fast and easy to use, with fully transparent results of good quality.

Broccoli supports a subset of SPARQL\footnote{\url{http://www.w3.org/TR/rdf-sparql-query}} (essentially trees with a single free variable at the root) for the ontology part of queries.
Moreover, it allows a special \emph{occurs-with} relation that can be used to specify co-occurrence of a class (e.g., \emph{plant}) or instance (e.g., \emph{Broccoli})
with an arbitrary combination of words, instances, and further subqueries.
Both traditional full-text search and pure ontology search are subsumed as special cases.
This gives a very powerful query language.
See Section \ref{sec:queries} for details.

For the \emph{occurs-with} relation, we provide a novel kind of pre-processing that decomposes sentences into \emph{contexts} of words that belong together.
In particular, this considers enumerations and {\subclauses}.
Previous approaches have used co-occurrence in a whole paragraph or sentence, or based on word proximity; all of these often give poor results.
See Section \ref{sec:preproc} for details.

We present the key idea behind a novel kind of index that supports fully interactive query times of around $100$ milliseconds and less for a collection as large as the full English Wikipedia (40 GB XML dump, 418 million contexts of the kind just described).
Previous approaches, including adaptations of the classic inverted index, yield query times on the order of seconds or even minutes for the kind of queries we support on collections of this size.
See Section \ref{sec:related:ontofull} for related work, and Section \ref{sec:index} for details.

All the described features have been implemented into a fully functional system with a comfortable user interface.
There is a single search field, as in full-text search, and suggestions are made after each keystroke.
This allows the user to incrementally construct semantic full-text queries without prior knowledge of a query language.
Results are ranked by relevance and grouped by instance, and displayed together with context snippets that provide full evidence for why that particular instance is shown.
See Figures \ref{figure:screenshot-1} and \ref{figure:screenshot-2} for an example, and Section \ref{sec:ui} for details.

We provide experimental results on the result quality for the English Wikipedia combined with the YAGO ontology \cite{DBLP:journals/ws/SuchanekKW08}.
For the quality results, we used 46 Queries from the SemSearch List Search Track (e.g., \emph{Apollo astronauts who walked on the Moon}), 15 queries from the TREC 2009 Entity Track benchmarks (e.g., \emph{Airlines that currently use Boeing 747 planes}) and 10 lists from Wikipedia (e.g. \emph{List of participating nations at the Winter Olympic Games}).
%We achieve very good results, and we even find many additional instances that were missing from the ground truth.
We allow reproducing our results at \url{http://broccoli.informatik.uni-freiburg.de/repro-corr/} .
%\url{http://vldb2013-506.hopto.org} .
%We find many instances that are missing from the manually compiled Wikipedia lists, although there is clear 
% in Wikipedia that they belong to the list.
See Section \ref{sec:experiments} for the details of our experiments.

We want to remark that the natural language processing, the index, and the user interface behind Broccoli are complex problems each on their own.
The contribution of this paper is the overall design of the system, the basic ideas for each of the mentioned components, an implementation of a fully functional prototype based on these ideas,
and a first performance and quality evaluation providing a proof of concept.
Optimization of the various components is the next step in this line of research; see Section \ref{sec:conclusions}.

%[Extensions of full-text search: error-tolerance, synonyms, etc.: useful but
%does not address real semantic queries. Ontology search: limited content and
%complex query language. Question answering: the ultimate goal, works fine for
%simple question but quickly becomes fuzzy / unusable when the questions become
%only slightly more complex / are not of the type foreseen by the builders of the
%search engine; Wolfra Alpha is a good example for that.]
%
%[Motivation behind our semantic full-text search: go a significant (semantic)
%step beyond full-text search, towards the ultimate goal of question
%answering, but stay within a limited and well-defined framework where things
%work perfectly and transparently. Refer to related work section to explanation of
%difference of our approach to question answering.]
%
%[Four components, briefly describe here and give forward pointer to the respective
%four sections in this paper: query language, index for fast query processing and
%suggestion, natural language processing, user interface.]
%
%[Brief summary of contribution, including experimental results.]
%
%[Absolutely put a screenshot in the paper, as early as possible. The screenshot
%should contain a carefully chosen example which highlights all the main features
%in a single picture. Take the one from Elmar's proposal as inspiration. A
%reference to SIGIR would be great.] 

%%prevent a widow
%\vspace{-2mm}

\section{Related work}\label{sec:related}

Putting the work presented in this paper into context is hard for two reasons.
First, the literature on semantic search technologies is vast.
Second, ``semantic'' means so many different things to different researchers.
We roughly divide work in this broad area into four categories, and discuss each category separately in the following four subsections.

\subsection{Combined ontology and full-text search}\label{sec:related:ontofull}

Ester \cite{DBLP:conf/sigir/BastCSW07} was the first system to offer efficient combined full-text and ontology search on a collection as large as the English Wikipedia.
Broccoli improves upon Ester in three important aspects.
First, Ester works with inverted lists for classes and achieves fast query times only on relatively simple queries. %; this is explained in more detail in Section \ref{sec:index:inverted}.
Second, Ester does not consider contexts but merely syntactic proximity of words / entities.
Third, Ester's simplistic user interface was ok for queries with one relation, but practically unusable for more complex queries.
%(1) Broccoli features a completely \emph{new index}, tailored for semantic full-text search, with consistently \emph{fast query times}; see Section \ref{sec:index}.
%Ester achieves subsecond query times only on relatively simple queries; see \cite[Section 7.1]{DBLP:conf/sigir/BastCSW07}
%(2) Broccoli does complex natural language processing in order to determine which words belong to the same context; see Section \ref{sec:preproc:nlp}.
%Ester approximates this issue with simple proximity search, which gives poor result quality for many queries; see \cite[Section 8]{DBLP:conf/sigir/BastCSW07}.
%(3) User interface.
%Ester offers a simplistic user interface with a single text field containing the whole query.
%This is ok for queries with one relation, but practically unusable for more complex queries.
%Broccoli offers a completely new user interface that allows the incremental construction of arbitray semantic full-text queries, intuitively guided by context-senstive suggestions after every keystroke; see Figures \ref{figure:screenshot-1} and \ref{figure:screenshot-2} and Section \ref{sec:ui}.

Various other systems offering combinations of full-text and ontology search have been proposed.
%
% Semplore. Queries like ours! Lucene. Abstract per entity. No NLP.
Semplore \cite{DBLP:journals/ws/WangLPFZTYP09} supports a query language similar to ours.
However, elements from the ontology are not recognized in their contexts, but there is simply one piece of text associated with each instance (which would correspond to a single large context in our setting).
Queries are processed with a standard inverted index %(see Section \ref{sec:index:inverted})
, and no particular UI is offered.
%
% Hybrid Search, ESWC 2008. Lucene + Sesame. No NLP. UI = searach field + ontology search mask.
In Hybrid Search \cite{DBLP:conf/esws/BhagdevCCLP08}, the full text and the ontology are searched separately with standard methods (Lucene and Sesame), and then the results are combined.
% Note: they are combined by showing documents containing triples from the ontology search together triples contained in documents from the full-text search.
There is no particular natural language processing.
%
% Concept Search, ESWC 2009. Build on Lucene. Yes NLP. No UI (keyword queries interpreted smtcly) 
Concept Search \cite{DBLP:conf/esws/GiunchigliaKZ09} adds information about identified noun phrases and hyponyms to the index.
Queries are bags of words, which are interpreted semantically.
The query processing uses standard methods (Lucene), with very long inverted lists for the semantic index items.
%
% GonNTogle, ESWC 2010. Lucene + annotation index. No Entity recog (manual annotations). No NLP.
GoNTogle \cite{DBLP:conf/esws/GiannopoulosBDS10} combines full text with annotations which are searched separately and then combined, similarly as in \cite{DBLP:conf/esws/BhagdevCCLP08}.
Queries are bags of words.
There is no full ontology search and no particular natural language processing.
%
% Neofonie Faceted Wikipedia Search. Abstract for each entity. Slow (DB). Nice UI.
Faceted Wikipedia Search \cite{DBLP:conf/bis/HahnBSHRBDS10} offers a user interface with similarities to ours.
However, the query language is restricted, there is nothing comparable to our contexts but only a small abstract per entity like in \cite{DBLP:journals/ws/WangLPFZTYP09}, and query processing is DB-based and very slow, despite the relatively small amount of data.
% 
% SIREN (from Sindice). Ontolgy search in Lucene. A framework, no more, no less.
SIREN\footnote{\url{http://siren.sindice.com}} provides an integration of pure ontology search into Lucene.
How to combine the then possible full-text and ontology searches is up to the user of the framework.
% From keywords to semantic, JWS 2009. Rewrites queries to SPARQL = pure ontology. No NLP. Yes UI.
Finally, systems like \cite{DBLP:journals/ws/ZenzZMSN09} try to interpret a given keyword query semantically and translate it into a suitable SPARQL query for pure ontology search.
%
% LEXXE. Single class + full text, very simplistic (synonym search). I found no paper so I won't cite it.

%\TODO{What about Sindice's SIREN = Semantic Information Retrieval Engine. See \url{http://siren.sindice.com}. In particular, you find there a short tutorial with some good examples of the type of contents they index and what kind of search they support. My understanding is that it is a pure ontology search engine. Which is integrated with Lucene though. Meaning that you can use Lucene for either. But they are not really combined. At least I do not see how and there is no mentioning of this anywhere.}

%However, all of these integrate full-text search in a very simplistic way only: each entity may have an associated text field and the search can be restricted to entities that co-occur with certain words in that text field.
%In contrast, Broccoli associates each single occurrence of an entity throughout the full text with a separate context.
%For example \emph{rhubarb stalks are edible} and \emph{rhubarb leaves are toxic} are two separate contexts, and better be, since otherwise we would get rhubarb as a hit when looking for plants with edible leaves; see Section \ref{sec:nlp}.
%Separate contexts for each entity occurrence poses a major efficiency challenge.
%Efficiency was not investigated in \TODO{Neofonie and Contentus and what else?}.

\newpage

\subsection{Systems for entity retrieval}\label{sec:related:entity}

\emph{Entity retrieval} is a line of research which focuses on search requests and corresponding result lists centered around entities (instead of around documents, as in traditional search).
Since 2009, there is also a corresponding Entity Track at TREC\footnote{\url{http://ilps.science.uva.nl/trec-entity}}.
The tasks of this track are both simpler and harder than what we aim at in this paper.

They are harder because the overall goal is entity retrieval from \emph{web pages}.
The ClueWeb09 collection introduced at TREC 2009 is 25 TB of text.
The relative information content is, however, low as is typical for web contents.
%\TODO{What about BTC and Sindince, can a similar thing be said about them? Hannah says: YES, they are also full of Muell.}
Moreover, identifying a representative web page for an entity is part of the problem.

To make the tasks feasible at all under these circumstances, the queries are relatively simple. 
For example, \emph{Airlines that currently use Boeing 747 planes}.%
\footnote{In our framework these are queries with two nodes and one \emph{occurs-with} edge.}
%Moreover, the entity types are restricted to one of: person, organization, location, product.
Even then the tasks remain very hard, and, for example, \emph{NDCG@R} figures average only around 30\% even for the best systems \cite{DBLP:conf/trec/BalogVSTW09}.

Broccoli queries can be trees of arbitrary degree and depth.
All entities that have a Wikipedia page are supported.
And, most importantly, the query process is interactive, providing the user with \emph{instant} feedback of what is in the collection and why a particular result appears.
This is key for constructing queries that give results of high quality.

The price we pay is a more extensive pre-processing assuming a certain ``cleanliness'' of the input collection.
Our natural language processing currently requires around 1600 core hours
%\footnote{And in the course of this research we did many such parses, each costing around US\$ 300 using Amazon EC2.}
on the 40 GB XML dump of the English Wikipedia. %, see Section \ref{sec:experiments:preproc}.
And Wikipedia's rule of linking the first occurrence of an important entity in an article to the respective Wikipedia article helps us for an entity recognition of good quality; see Section \ref{sec:preproc:er}.
Bringing Broccoli's functionality to web search is a very reasonable next step, but out of scope for this article.

Another popular form of entity retrieval is known as \emph{ad-hoc object retrieval} \cite{DBLP:conf/www/PoundMZ10}.
Here, the search is on structured data, as discussed in the next subsection.
Queries are given by a sequence of keywords, similar as in full-text search, for example, \emph{doctors in barcelona}.
Then query interpretation becomes a non-trivial problem; see Section \ref{sec:related:qa}.

\subsection{Information extraction and ontology search}\label{sec:related:ontoonly}

Systems for ontology search have reached a high level of sophistication.
For example, RDF-3X can answer complex SPARQL queries on the Barton dataset (50 million triples) in less than a second on average \cite{DBLP:journals/vldb/NeumannW10}.

As part of the Semantic Web / Linked Open Data \cite{DBLP:journals/ijswis/BizerHB09} effort, more and more data is explicitly available as fact triples.
The bulk of useful triple data is still harvested from text documents though. 
The information extraction techniques employed range from simple parsing of structured information (for example, many of the relations in YAGO or DBpedia \cite{DBLP:conf/semweb/AuerBKLCI07} come from the Wikipedia info boxes) over pattern matching (e.g., \cite{DBLP:conf/dl/AgichteinG00}) to complex techniques involving non-trivial natural language processing like in our paper (e.g., \cite{DBLP:conf/ijcai/BankoCSBE07}). For a relatively recent survey, see \cite{DBLP:journals/ftdb/Sarawagi08}.

Our work differs from this line of research in two important aspects: (1) the full text remains part of the index that is searched at query time; and (2) our system is fully interactive and keeps the human in the loop in the information extraction process.
This has the following advantage:

Ontologies are good for facts like \emph{which plants are native to which regions}, \emph{who was born where on which date}, etc.
Such facts are easy to define and can be extracted from existing data sources in large quantity and with reasonable quality.
And once in the ontology, they are easily combinable, permitting queries that would not work with full-text search.

But for more complex facts like our \emph{broccoli has edible leaves}, it is the other way round.
They are easy to express and search in full text, but tedious to define, include, and maintain in an ontology.
Let alone the problem of guessing the right relation names when searching for them.

By keeping the full text, we can leverage the intelligence of the user at query time.
The query \emph{Plant occurs-with edible leaves} does not specify the type of the relation between the occurrence of the plant and the occurrence of the words \emph{edible} and \emph{leaves}.
Yet a moment's thought reveals that it is quite likely that a context matching these elements gives us what we want.
Similarly as in full-text search, there is often no need to be overly precise in order to get what you want.
And just like the result snippets in full-text search, Broccoli's result snippets provide instant feedback on whether the listed plant is really one with edible leaves.

Finally, if information extraction is desired nevertheless, Broccoli can be a useful tool for interactively exploring the collection with respect to the desired information, and for formulating appropriate queries.

%Finally, one could argue that whatever we do here could just as well be used to extract fact triples from the full text, for subsequent retrieval via a pure ontology search engine.
%Indeed, we consider this as one important application of our work.
%However, the difference is that with our system the information extraction happens at query time.

%However, we see a big advantage in keeping the full text until query time.
%Namely, one reason for the success of full-text search engines is that the service they provide is, in principle, very basic and transparent: you get documents containing some or all of your keywords, no more, no less.
%The actual intelligence comes from the user, via the choice of keywords.
%Our semantic full-text search leverages user intelligence in a very similar way.

%and providing mappings between namings in different ontologies.
%\footnote{%
%The two separate triples \emph{Broccoli has-part leaf} and \emph{leaf has-property edible} would lack the information that not all leaves are edible.
%One conceivable fix would be the introduction of an entity \emph{edible leaf}.
%Another conceivable fix would be to introduce the unary relation \emph{has-edible-leaf}.
%Both fixes are hard to automatize, and manually maintaining entity and relation names of such specificity seems infeasible.
%Let alone the usability problem of finding the appropriate entity and relation name at query time.}.

\subsection{Systems for question answering}\label{sec:related:qa}

Question answering (QA) systems provide similar functionality as our semantic full-text search.
The crucial difference is that questions can be asked in natural language, which makes the answering part much harder.
The system is burdened with the additional and very complex task of ``translating'', in one way or the other, the given natural language query into a more formal query or queries that can be fed to a search engine and / or a knowledge database.

The perfect QA system would obviate the need for a system like ours here.
But research is still far from achieving that goal.
All state-of-the-art QA systems, including the big commercial ones, are specialized to quite particular kinds of questions.
For example, Wolfram Alpha works perfectly for \emph{Which cities in China have more than 10 million inhabitants}, but does not work if \emph{more} is replaced by \emph{less} or \emph{China} by \emph{Asia}, and does not even understand the question \emph{Which plants have edible leaves}.
IBM's Watson was tuned for finding the single most probable entity when given one of the (intentionally obscured) clues from the {Jeopardy!} game.
And both of these systems lack transparency: it is hard to predict whether a question will be understood correctly, it is hard to understand the reasons for a missing or wrong answer, and
there is no possibility of interaction or query refinement.

For our semantic full-text search both the query language and the relation between a given query and its result are well-defined and maximally transparent to the user; see the discussion in Section \ref{sec:related:ontoonly}.
The price we pay is query formulation in a non-natural language.
%, just like in ordinary full-text search.
The success of full-text search has shown that as long as the language is simple enough, it can work.
%Given the current state of the art and the success of full-text search, we consider this a very reasonable compromise.

\newpage

\section{Input data and natural language pre-processing}\label{sec:preproc}

\subsection{Input data}\label{sec:preproc:input}

Broccoli requires two kinds of inputs, a text collection and an ontology.
The text collection consists of documents containing plain text.
The ontology consists of typed \emph{relations} with each relation containing an arbitrary set of fact triples.
The subjects and objects of the triples are called \emph{instances}.
Each instance belongs to one or more \emph{classes}.
The classes are organized in a taxonomy; the root class is called \emph{Entity}.

\subsection{Entity recognition}\label{sec:preproc:er}

The first step is to identify mentions of or referrals to instances from the ontology in the text documents.
Consider the following sentence, which will be our running example for this section:

(S) \emph{The usable parts of rhubarb, a plant from the Polygonaceae family, are the medicinally used roots and the edible stalks, however its leaves are toxic.}

Both \emph{rhubarb} and \emph{its} refer to the instance \emph{Rhubarb} from our ontology, which in turn belongs to the classes \emph{Plant} and \emph{Vegetable} (among others).

Our entity recognition on the English Wikipedia is simplistic but reasonably effective.
As a rule, first occurrences of entities in Wikipedia documents are linked to their Wikipedia page.
When parsing a document, whenever a part or the full name of that entity is mentioned again in the same section of the document (for example, \emph{Einstein} referring to \emph{Albert Einstein}), we recognize it as that entity.

We resolve anaphora in an equally simplistic way. Namely, we assign each occurrence of \emph{he}, \emph{she}, \emph{it}, \emph{her}, \emph{his}, etc.\ to the last recognized entity of matching gender.
%NOTE(Florian): removed the footnote from "gender" to save space
%\footnote{The YAGO ontology does not provide enough information to tell male from female, but we can easily tell male/female from neuter.}.
We also recognize the pattern \emph{the $<$class$>$} as the entity of the document if it belongs to \emph{$<$class$>$}, for example, \emph{the plant} in the document of \emph{Broccoli}.

Our results in Section \ref{sec:experiments:quality} suggest that, on Wikipedia, these simple procedures give already a reasonable accuracy.

\subsection{Natural language processing}\label{sec:preproc:nlp}

The second step is to decompose document texts into what we call \emph{contexts}, that is, sets of words that ``belong'' together.
The contexts for our example sentence (S) from above are:

\smallskip
\noindent (C1) \emph{\underline{rhubarb}, a plant from the Polygonaceae family}
\par\smallskip
\noindent (C2) \emph{The usable parts of \underline{rhubarb} are the medicinally used roots}
\par\smallskip
\noindent (C3) \emph{The usable parts of \underline{rhubarb} are the edible stalks}
\par\smallskip
\noindent (C4) \emph{however \underline{rhubarb} leaves are toxic}
\smallskip

\noindent
This will be crucial for the quality of our results, because we do not want to get rhubarb in our answer set when searching for \emph{plants with edible leaves}.
Note that we assume here that the entity recognition and anaphora resolution have already been done (underlined words).
Also note that we do not care whether our contexts are grammatically correct and form a readable text.
This distinguishes our approach from a line of research called \emph{text simplification} \cite{DBLP:conf/coling/ChandrasekarDS96}.
%\cite{DBLP:conf/coopis/KlebanovKM04}.
% NOTE(Hannah): Klebanov wants grammatically correct sentences because there the sentence simplification comes *before* the parsing, and is supposed to make the work of the parser simpler. We do it the other way round. And in the long run, we want to do it on our own, without any full parse needed before or after.

% In the following, as in the example above, we will only consider contexts that are part of a single sentence.
In the following, we will only consider contexts that are part of a single sentence.
Indeed, after anaphora resolution, it seems that most simple facts are expressed within one and the same sentence.
Our evaluation in Section \ref{sec:experiments:quality} confirms this assumption.
% The results from our quality evaluation in Section \ref{sec:experiments:quality} (error category FN2 in Table \ref{table:error-categories}) confirm this assumption.

Our context decomposition consists of two parts, each described in the following subsections.

\subsubsection{Sentence constituent identification (SCI)}\label{sec:preproc:sci}

The task of SCI is to identify the basic ``building blocks'' of a given sentence.
For our purposes various kinds of \emph{\subclauses} and \emph{enumeration items}
will be important, because they usually contain separate facts that have no direct relationship to the other parts of the sentence.
For example, in our sentence (S) from above, the relative clause \emph{a plant from the Polygonaceae family} refers to \emph{rhubarb} but has nothing to do with the rest of the sentence.
Similarly, the two enumeration items \emph{the medicinally used roots} and \emph{the edible stalks} have nothing to do with each other (except that they both refer to rhubarb); in particular, rhubarb roots are not edible and rhubarb stalks are not medicinally used.
Finally the part \emph{however its leaves are toxic} needs to be considered separate from the preceding part of the sentence.
As will become clear in the following, we consider these as enumeration items on the top level of the sentence.

% NOTE(Hannah): this paragraph could be commented out in case we need space. SSS
% In general, {\subclauses} can again contain enumerations, some or all of which can again contain {\subclauses}, and so on.
% And vice versa.
% In practice the nesting level is one or two for most sentences.

\begin{figure}[ht]
\vspace{-2mm}
\hspace{-4.5mm}
\includegraphics[width=115mm]{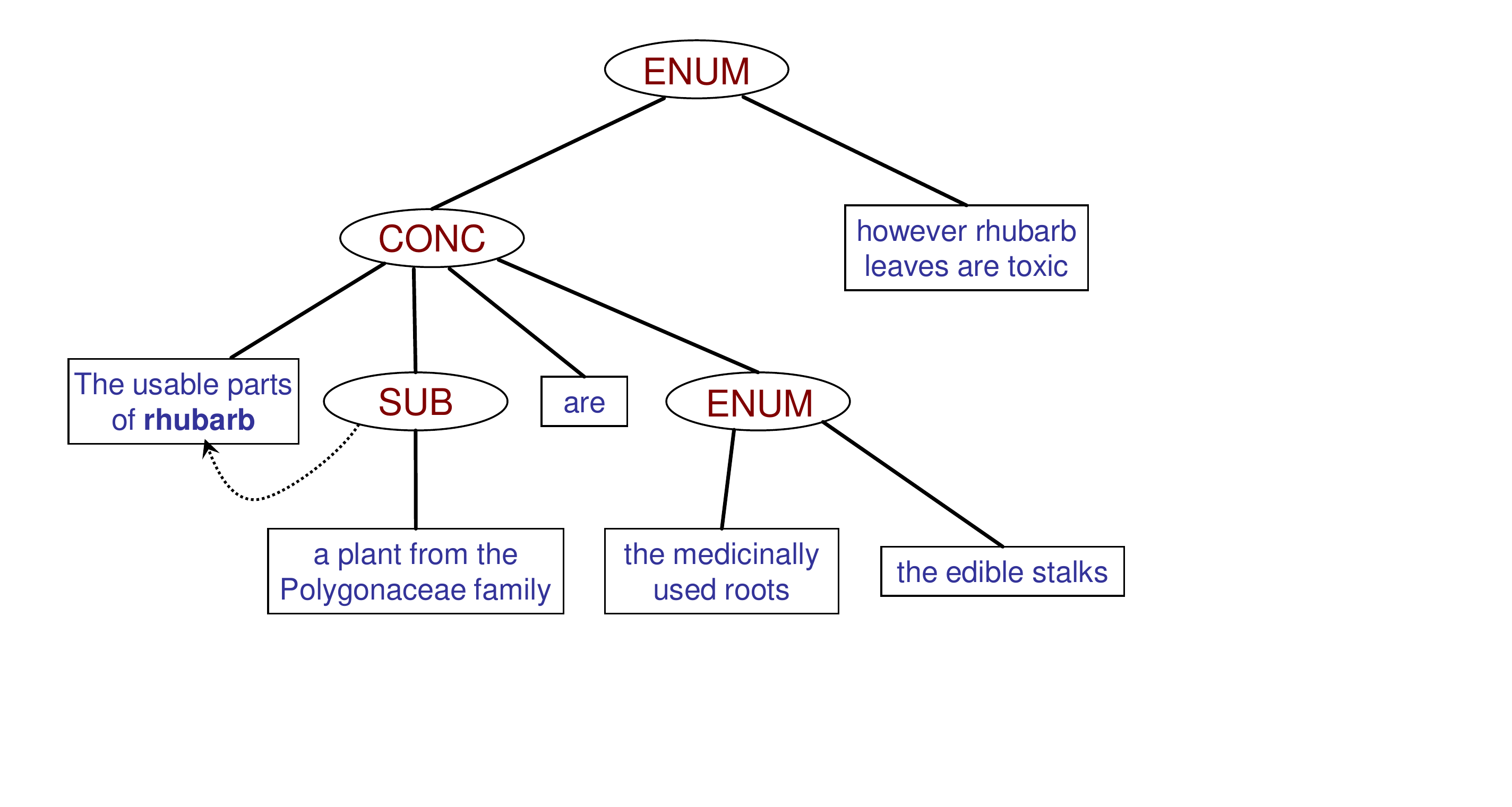}
\vspace{-16mm}
\vspace{-3mm} %NOTE(Florian): further reduce the space between the figure and its caption for the sake of space
\caption{The SCI tree for our example sentence after anaphora resolution.
The head of the {\subclause} is printed in bold.%
\label{figure:sci-example}}
\end{figure}

Formally, SCI computes a tree with three kinds of nodes: \emph{enumeration (ENUM)}, \emph{{\subclause} (\SUB)}, and \emph{concatenation (CONC)}.
The leaves contain parts of the sentence and a concatenation of the leaves from left to right yields the whole sentence again.
See Figure \ref{figure:sci-example} for the SCI tree of the above sentence.

We construct our SCI trees based on the output of a state-of-the-art constituent parser.
We use SENNA \cite{DBLP:journals/jmlr/Collobert11}, because of its good trade-off between parse time (around 35ms per sentence) and result quality (see Section \ref{sec:experiments:quality}).

We transform the parse tree using a relatively small set of hand-crafted rules.
Here is a selection of the most important rules; the complete list consists of only 11 rules but is omitted here for the sake of brevity.
In the following description when we speak of an \emph{NP} (noun phrase), \emph{VP} (verb phrase), \emph{SBAR} (subordinate clause), or \emph{PP} (prepositional phrase)  we refer to nodes in the parse tree with that tag.
% and sometimes also to the part of the sentence under that node.

% NOTE(Hannah): I have left out S from the list here because this tag lacks a clear specification (all kinds of "sub-sentences" are S) and we deal with it just like with an SBAR without head. 
\medskip\noindent
(SCI 1) Mark as ENUM each node, for which the children (excluding punctuation and conjunctions) are either all \emph{NP} or all \emph{VP}.
\par\smallskip\noindent
(SCI 2) Mark as {\SUB} each \emph{SBAR}.
If it starts with a word from a positive-list (e.g., \emph{which} or \emph{who}) define the first \emph{NP} on the left as the \emph{head} of this {\SUB}; this will be used in (SCR 0) below.
\par\smallskip\noindent
(SCI 3) Mark as {\SUB} each \emph{PP} starting with a preposition from a positive-list (e.g., \emph{before} or \emph{while}), and all \emph{PP}s at the beginning of a sentence. These {\SUB}s have no head.
\par\smallskip\noindent
(SCI 4) Mark as CONC all remaining nodes and contract away each CONC with only text nodes in its subtree (by merging the respective text).
\medskip

As our quality evaluation in Section \ref{sec:experiments:quality} shows, our rules work reasonably well.

\subsubsection{Sentence constituent recombination (SCR)}\label{sec:preproc:scr}

In SCR we recombine the constituents identified by the SCI to form our \emph{contexts}, which will be the units for our search.
Recall that the intuition is to have contexts such that only those words which ``belong'' together are in the same context.
SCR recursively computes the following contexts from a SCI tree or subtree:

\medskip\noindent
(SCR 0) Take out each subtree labeled {\SUB}.
If a head was defined for it in (SCI 2), add that head as the leftmost child (but leave it in the SCI tree, too).
Then process each such subtree and the remaining part of the original SCI tree
(each of which then only has ENUM and CONC nodes left) separately as follows:
\par\medskip\noindent
(SCR 1) For a leaf, there is exactly one context: the part of the sentence stored in that leaf.
\par\smallskip\noindent
(SCR 2a) For an inner node, first recursively compute the set of contexts for each of its children.
\par\smallskip\noindent
(SCR 2b) If the node is marked ENUM, the set of contexts for this node is computed as the \emph{union} of the sets of contexts of the children.
\par\smallskip\noindent
(SCR 2c) If the node is marked CONC, the set of contexts for this node is computed as the \emph{cross-product} of the sets of contexts of the children.
\medskip

We remark that once we have the SCI tree, SCR is straightforward, and that the time for both SCI + SCR is negligible compared to the time needed for the full-parse of the sentences.

%prevent an orphan
%\vspace{1mm}

\section{Query language}\label{sec:queries}

%We briefly explain what kinds of queries Broccoli can process, and what constitutes an answer to a given such query.
%
%\subsection{Queries}

Queries to Broccoli are rooted trees with arcs directed away from the root.
The root is either a class or an instance.
There are two types of arcs: \emph{ontology arcs} and \emph{occurs-with arcs}.
Both have a class or instance as source node.%
%\footnote{For queries with an instance at the root, at most one result is returned: that instance (if the rest of the query matches). These queries are still useful for browsing the evidence for that particular hit.}

Ontology arcs are labeled by a relation from the ontology.
The two nodes must be classes or instances matching the source and target type of the relation.
The class or instance at the target node may be the root of another arbitrary tree.

For occurs-with arcs, the target node can be an arbitrary set of words, prefixes, instances or classes.
The instances or classes may themselves be the root of another arbitrary query.
Example queries are given in Figures \ref{figure:screenshot-1} and \ref{figure:screenshot-2}.

% Note(Björn): If spance is needed, deactive the next paragraph maybe
To give an example of a more complex query: in Figure \ref{figure:screenshot-1} we could replace the instance node \emph{Europe} by a class node \emph{Location} and add to it an \emph{occurs-with} arc with the word \emph{equator} in its target node.
The intention of this query would be to obtain plants with edible leaves native to regions at or near the equator.

%For our definition of the answer set for a given query, we refer to Section \ref{sec:index:queryproc}, which describes the (recursive) algorithm for processing an arbitrary given query.
%The query processed internally is similar with one exception:
%Instances and classes are replaced by a variable and an additional arc with an \emph{is-a} relation (or \emph{equals}, respectively).

\vfill

%\subsection{Answers}
%
%The answer to a query is a set of contexts grouped by instance.
%Both the instances and the contexts within an instance are ranked, we come back to that \TODO{where?}.
%For a given query we recursively define the answer as follows.
%
%(1) The answer for a single \emph{word} or \emph{prefix} is a subset of contexts mentioning that word or prefix, grouped by the document they belong to. \TODO{Do we really have/need a 1-1 correspondence between documents and instances? Can't we also have documents that belong to no instance? Then again, we could just define an instance for each document if we wanted to.}
%The answer for a single \emph{instance} is a subset of contexts mentioning that instance, in a single group.
%The answer for a single \emph{class} is a subset of contexts containing instances from that class, grouped by instance.
%
%(2) The answer for a node with children is as follows.
%\TODO{Explain 1: what is the answer for each child. Explain 2: what to do for each ontological node and for each occurs-with arc. Explain 3: how to combine these.}

\section{Index and query processing}\label{sec:index}

The index and query processing of Broccoli are described in detail in \cite{broindex}.
In this section, we summarize why standard indexes are not suited for Broccoli and describe the main idea behind our new index.

There are sophisticated systems for both, full-text search and search in ontologies.
Since our queries combine both tasks, three ways to answer our queries using those system come to mind:
(1) incorporate ontology information into an inverted index; 
(2) incorporate full-text information into a triple store; 
(3) use an inverted index for the full-text part of the query, a triple store for the ontology part of the query, and then combine the results somehow.

Neither approach is perfectly suited for our use-case. 
In a nutshell, approach (1) produces document-centric results and cannot be used to answer complex queries that involve join operations.
Approach (2) needs a relation (e.g. \emph{occurs-in-context} featuring both, words and entities) of the size of our entire index to make use of the contexts produced in our contextual sentence decomposition. 
Efficient queries require a special purpose index over this relation, which already goes in the direction of our approach.
Finally, approach (3) will get a list of contexts as a result from the full-text index and has to derive all entities that occur in those contexts. 
This mapping is not trivial to achieve efficiently, especially since a full mapping from contexts to entities usually does not fit in memory for large collections.
Apart from that, we allow queries that demand co-occurrence with some entity from a list that can be the root of another query (e.g. a query for politicians that are friends with an astronaut who walked on the moon). 
This would require a second mapping in the other direction: from entities to contexts. 
In summary, the two problems are: 
Given a list of contexts $C$, produce a list $E$ of entities that occur in those contexts. 
Given a list of contexts $C$ and an entity list $E$, limit $C$ to contexts that include at least one entity from $E$. 

The main idea behind our new index solves these two problems. 
We use what we call \emph{context lists} instead of standard inverted lists.
The context list for a prefix contains one index item per occurrence of a word starting with that prefix, just like the inverted list for that prefix would.
But along with that it also contains one index item for each occurrence of an arbitrary entity in the same context as one of these words.
For example, consider the context \emph{the usable parts of \underline{rhubarb} are its edible \underline{stalks}}, with recognized entities underlined.
And let us assume that we have an inverted list for each 4-letter prefix.
Then the part of the context list for \emph{edib*} pertaining to this context (which has id, say, 14) would be:

\medskip
{\renewcommand{\baselinestretch}{1.3}\normalsize
\hspace{2mm}
\begin{tabular}{cccccc}
\multirow{4}{*}{\hspace{-3mm}edib*:\hspace{2mm}}
& ... & C14 & C14 & C14 & ... \\
& ... & \#edible & \#Rhubarb & \#Stalk & ... \\
& ... &        1 &         1 &       1 & ... \\
& ... &        8 &         5 &       9 &  ... \\
\end{tabular}}
\medskip

\noindent
The numbers in the first row are context ids.
The \# in the second row means that not the actual entities (with capital letters) or words are stored, but rather unique ids for them.
The third row contains the score for each index item.
The fourth row contains the position of the word or entity in the respective context.
The context lists are sorted by context id, and, for equal context ids, by word/entity id, with entities coming after the words.

Since entity postings are included in those lists, we can easily solve the two problems introduced above.
Actually, our index and query processing support many additional features like excerpt generation, suggestions, prefix search, search for documents instead of entities or ranges over values. 
For details on those features and a detailed description of the query processing, we again refer the reader to \cite{broindex}.

\section{User interface}\label{sec:ui}

For a convincing proof of concept for our interactive semantic search, we have taken great care to implement a fully functional and intuitive user interface.
In particular, there is no need for the user to formulate queries in a language like SPARQL.
% NOTE(Björn): Outcommented in search of more space:
% NOTE(Florian): Added this again because of lots of available space:
We claim that any user familiar with full-text search will learn how to use Broccoli in a short time, simply by typing a few queries and following the various query suggestions.
% NOTE(Björn): This was already outcommented:
% NOTE(Florian): Added this again because of lots of available space:
The user interface is completely written in Java using the Goole Web Toolkit\footnote{\url{http://code.google.com/webtoolkit}}.

The introduction and screenshots (Figures \ref{figure:screenshot-1} and \ref{figure:screenshot-2}) have already provided a foretaste of the capabilities of our user interface.
%Here is a list of further features.
%Due to the space constraint, we mention only the most important ones:
Here is a list of its most important further features:

\medskip\noindent
(UI 1) Search as you type: New suggestions and results with every keystroke.
Very importantly, Broccoli's suggestions for words, classes, instances, and relations are context-sensitive.
That is, the displayed suggestions actually lead to hits, and the more / higher-scored hits they lead to, the higher they are ranked.

\smallskip\noindent
(UI 2) Pre-select of most likely suggestion: Broccoli knows four kinds of objects: words, classes, instances, and relations.
Depending on where you are in the query construction, you get suggestions for several of them.
A new user may be overwhelmed to understand the different semantics of the different boxes.
For that reason, after every keystroke Broccoli highlights the most meaningful suggestion, which can be selected by simply pressing \emph{Return}.
%NOTE(Florian): removed the following addition to the previous sentence because the caption does noe explain much about that and to sace space
%; see Figure \ref{figure:screenshot-2} and the explanation in the caption. 

\smallskip\noindent
(UI 3) Visual query representation: At any time, the current query is shown as a tree, with a
color code for the various elements that is consistent with the suggestion boxes.
%NOTE(Florian): removed the following addition to the previous sentence to save some space (and because it is redundant because the screenshots are already referenced)
%; see Figures \ref{figure:screenshot-1} and \ref{figure:screenshot-2}.

\smallskip\noindent
(UI 4) Change of focus / root: A click on any node in the query tree will change the focus of the query suggestions to that node.
A double-click on any class or instance node will make that node the root of the tree and re-group and re-rank the results accordingly.

\smallskip\noindent
(UI 5) Full history support: The forward and backward buttons of the browser can be used to undo or redo single steps of the query creation process.
Furthermore the current URL of the interface can always be used to store its current state or to exchange created queries with others.

\smallskip\noindent
(UI 6) Tutorial: Besides some pre-built example queries, the interface also provides a tutorial mode that shows how to create a search query step by step.

%\medskip
%We encourage the reader to have a look at the user interface in action by trying our online demo that is accessible via \url{http://vldb2013-506.hopto.org}.

%\vspace{10mm}

\section{Experiments}\label{sec:experiments}

\subsection{Input data}\label{sec:experiments:input}

Our text collection is the text from all documents in the English Wikipedia,
obtained via \url{download.wikimedia.org} in January 2013.
Some dimensions of this collection: 40 GB XML dump, 2.4 billion word
occurrences (1.6 billion without stop-words), 285 million recognized entity occurrences 
and 200 million sentences which we decompose into 418 million contexts.

As ontology we use the latest version of YAGO from October 2009.
We manually fixed 92 obvious mistakes in the ontology (for example, the \emph{noble prize} was a \emph{laureate} and hence a \emph{person}), 
%removed some classes and relations of poor quality,
and added the relation \emph{Plant native-in Location} for demonstration purposes.
Altogether our variant of YAGO contains 2.6 million entities, 19,124 classes, 60 relations, and
26.6 million facts.

% NOTE(Björn): Possibly remove to free up more space
%We remark again that, in principle, our approach works for \emph{any} given text collection, and for \emph{any} given ontology.
%It is just that we currently use a relatively simplistic \emph{entity recognition} tuned for Wikipedia; see Section \ref{sec:preproc:er}. 
%This could, however, be replaced by any other high-quality entity recognition.
\subsection{Pre-processing}\label{sec:experiments:pre-proc}
We use a UIMA\footnote{\url{http://uima.apache.org/}} pipeline to pre-process the Wikipedia XML.
The pipeline includes self-written components to parse the Wikipedia markup, tokenize text, parse sentences using SENNA \cite{DBLP:journals/jmlr/Collobert11}, perform entity-recognition and anaphora resolution (see section \ref{sec:preproc:er}), and decompose the sentences (see section \ref{sec:preproc:nlp}).
We want to note that all these components can easily be exchanged.
In principle, this allows Broccoli to work with any given text collection and ontology.

The full parse with SENNA was scaled out asynchronously on a cluster of 8 PCs, each equipped with an AMD FX-8150 8-core processor and 16 GB of main memory.
A final non-UIMA component writes the binary index which is kept in three separate files.
The file for the context lists has a size of 37 GB. 
The file for the relation lists has a size of 0.5 GB. 
And the file for the document excerpts has a size of 276 GB, which could easily be reduced to 85 GB by eliminating the redundant and debug information the file currently contains. 

\subsection{Computing environment}\label{sec:experiments:environment}

The code for the index building and query processing is written entirely in C++.
The code for the query evaluation is written in Perl, Java, C++ and JavaScript.
Our pre-processing components are written in C++ or Java.
% The full parse using SENNA (see Section \ref{sec:preproc:nlp}) was run in parallel on 20 high-CPU instances on Amazon EC2, at a cost of around \$US 200.
All performance tests were run on a single core of a Dell PowerEdge server with 2 Intel Xeon 2.6 GHz processors, 96 GB of main memory, and 6x900 GB SAS hard disks configured as Raid-5.

%\subsection{Pre-processing times and space usage}\label{sec:experiments:preproc}
%The parallelized full parse using SENNA takes around 19 hours.
%Basic parsing of the Wikipedia with our simple entity recognition, anaphora resolution and SCI + SCR takes around 6 hours.
%Going from there, building the index lists takes another 3 hours.
%
%\TODO{Remove or change to values from the other paper}
%Our index is kept in three separate files.
%The file for the context lists has a size of 42 GB.
%The total number of postings is 2.8 times as much as in a standard full-text index.
%The file for the relation lists has a size of 1 GB.
%The file for the document excerpts has a size of 73 GB, which could easily be reduced to 51 GB by eliminating the redundant information the file currently contains.
%Compression, which we currently do not use anywhere, could reduce the size of these files further.
%\TODO{Is the meaning of the sentence correct? The estimate is based on the sum of: 4-byte/doc-id, size of (unique) doc-url+titel, and the size of all excerpts. Hannah says: YES, sounds fine to me.}

\subsection{Query times}\label{sec:experiments:query-times}

For detailed experiments on query times, we refer to the paper describing the
index behind Broccoli \cite{broindex}.
In said paper, we have evaluated our system on 8,000 queries of different complexity and 35,000 suggestions. 
%For the sake of brevity, we omit a detailed breakdown and limit ourselves to
Therefore we here omit a detailed breakdown and limit ourselves to
the figures reported in Table \ref{table:query-times}.

% 8000 hit queries
% 35141 suggestions
\begin{table}[ht]
{\renewcommand{\baselinestretch}{1.3}\normalsize
\begin{tabular}{|l||c||c|c|c|}
\hline
Query set   &     average &       median &   $90\ptile$ &   $99\ptile$ \\\hline\hline
Hit queries &   $52\ms$ &   $ 23\ms$ & $139\ms$ & $393\ms$ \\
Suggestion  &   $19\ms$ &   $  6\ms$ &  $44\ms$ & $193\ms$ \\
\hline
\end{tabular}}
\caption{Statistics of query times over 8,000 queries and 35,000 suggestions.}
\label{table:query-times}
\end{table}

\def\FP{\makebox[10mm]{\#FP}}
\def\FN{\makebox[10mm]{\#FN}}
\def\prec{\makebox[10mm]{Precision}}
\def\rec{\makebox[10mm]{Recall}}
\def\f1{\makebox[10mm]{F1}}
\def\pten{\makebox[9mm]{P@10}}
\def\rprec{\makebox[9mm]{R-Prec}}
\def\map{\makebox[9mm]{MAP}}
\def\phun{\makebox[9mm]{P@100}}
\def\ndcgr{\makebox[9mm]{nDCG}}
\def\siga{$^{\dagger}$\!\!\!}
\def\sigb{$^{\ast}$\!\!\!}
% The table is defined here, so that it appears at the correct place. Might have to move around later.
\begin{table*}[!t]
\centering
%\hspace*{-1.5mm}
\begin{tabular}{|+l|^l||^c|^c|^c|^c|^c|^c|^c|^c|^c|}
\hline
                                  &             &       \FP  &     \FN  &   \prec  &    \rec  &     \f1       &   \pten  &  \rprec  &   \map   &  \ndcgr  \\\hline\hline
\multirow{3}{*}{SemSearch}        &  sections   &  $44,117$  &    $92$  &  $0.06$  &  $0.78$  &  $0.09$       &  $0.32$  &  $0.42$  &  $0.44$  &  $0.45$  \\
                                  &  sentences  &   $1,361$  &   $119$  &  $0.29$  &  $0.75$  &  $0.35$       &  $0.32$  &  $0.50$  &  $0.49$  &  $0.50$  \\
\rowstyle{\bfseries\boldmath}
                                  &  contexts   &     $676$  &   $139$  &  $0.39$  &  $0.67$  &  $0.43$\siga  &  $0.25$  &  $0.52$  &  $0.45$  &  $0.48$  \\\hline
\multirow{3}{*}{Wikipedia lists}  &  sections   &  $28,812$  &   $354$  &  $0.13$  &  $0.84$  &  $0.21$       &  $0.46$  &  $0.38$  &  $0.33$  &  $0.41$  \\
                                  &  sentences  &   $1,758$  &   $266$  &  $0.49$  &  $0.79$  &  $0.58$       &  $0.82$  &  $0.65$  &  $0.59$  &  $0.68$  \\
\rowstyle{\bfseries\boldmath}
                                  &  contexts   &     $931$  &   $392$  &  $0.61$  &  $0.73$  &  $0.64$\sigb  &  $0.84$  &  $0.70$  &  $0.57$  &  $0.69$  \\\hline
\multirow{3}{*}{TREC}             &  sections   &   $6,890$  &    $19$  &  $0.05$  &  $0.82$  &  $0.08$       &  $0.28$  &  $0.29$  &  $0.29$  &  $0.33$  \\
                                  &  sentences  &     $392$  &    $38$  &  $0.39$  &  $0.65$  &  $0.37$       &  $0.58$  &  $0.62$  &  $0.46$  &  $0.52$  \\
\rowstyle{\bfseries\boldmath}
                                  &  contexts   &     $297$  &    $36$  &  $0.45$  &  $0.67$  &  $0.46$\sigb  &  $0.58$  &  $0.62$  &  $0.46$  &  $0.55$  \\\hline
\end{tabular}
\vspace{2mm}
\caption{Sum of false-positives and false-negatives and averages for other measures over all SemSearch, Wikipedia list and TREC queries for Broccoli when running on sections, sentences or contexts.
For contexts, the results for the SemSearch and Wikipedia list benchmarks can be reproduced using our web application at \url{http://broccoli.informatik.uni-freiburg.de/repro-corr/} .
%\url{http://vldb2013-506.hopto.org} .
$\ast$, $\dagger$ denotes a p-value $<0.02$, $<0.003$ for the two-tailed t-test against the sentences baseline.
\label{table:quality-results}}
%\vspace{-1mm}
\end{table*}

On our collection, 90\% of the queries finish within 140ms, 99\% within 400ms. 
Suggestions are even faster. The breakdown in \cite{broindex} shows that for a
combination of Wikipedia and YAGO, only queries that include text take siginificant
time. Purely ontological queries finish within 2ms on average.

\subsection{Result quality}\label{sec:experiments:quality}

We performed an extensive quality evaluation using topics and relevance judgments from several standard benchmarking tasks for entity retrieval: the Yahoo SemSearch 2011 List Search Track \cite{DBLP:conf/www/TranMWG11}, the TREC 2009 Entity Track \cite{DBLP:conf/trec/BalogVSTW09} and, similarly as in \cite{DBLP:conf/sigir/BastCSW07}, a random selection of ten Wikipedia featured \emph{List of ...} pages.
To allow reproducability we provide queries and relevance judgments as well as the possibilty to evaluate (and modify) the queries against a live running system for the SemSearch List Track and the Wikipedia lists at \url{http://broccoli.informatik.uni-freiburg.de/repro-corr/} .
%\url{http://vldb2013-506.hopto.org} .
The TREC Entity Track queries were used for an in-depth quality evaluation that does not allow for an easy reproduction.
Therefore we do not provide them in our reproducability web application.
In the following we first describe each of the tasks in more detail.

The SemSearch 2011 List Search Track consisted of 50 topics asking for lists of entities in natural language, e.g. \emph{Apollo astronauts who walked on the Moon}.
The publicly available results were created by pooling the results of participating systems and are partly incomplete.
Furthermore, the task used a subset of the Billion Triple Challenge Linked Data as collection, and some of the results referenced the same entity several times, e.g. once in DBPedia and once in OpenCyc.
Therefore, we manually created a new ground truth consisting of Wikipedia entities.
This is possible because most topics were inspired by Wikipedia lists and can be answered completely by manual investigation.
Three of the topics did not contain any result entities in Wikipedia, and we ignored one additional topic because it was too controversial to answer with certainty (\emph{books of the Jewish canon}).
This leaves us with 46 topics and a total of 384 corresponding entities in our ground truth\footnote{\label{foot:repro-tool} available at \url{http://broccoli.informatik.uni-freiburg.de/repro-corr/}} .
%\url{http://vldb2013-506.hopto.org}} .
The original relevance judgments only had 42 topics with primary results and 454 corresponding entities, including many duplicates.

The TREC 2009 Entity Track worked with the ClueWeb09 collection and consisted of 20 topics also asking for lists of entities in natural language, e.g. \emph{Airlines that currently use Boeing 747 planes}, but in addition provided the source entity (\emph{Boeing 747}) and the type of the target entity (\emph{organization}).
We removed all relevance judgments for pages that were not contained in the English Wikipedia; this approach was taken before in \cite{DBLP:conf/cikm/BronBR10} as well.
This leaves us with 15 topics and a total of 140 corresponding relevance judgments.

As third benchmark we took a random selection of ten of Wikipedia's over 2,400 manually compiled featured \url{en.wikipedia.org/wiki/List_of_...} pages\footnote{\url{http://en.wikipedia.org/wiki/Wikipedia:Featured_lists}}, e.g. the \emph{List of participating nations at the Winter Olympic Games}.
Wiki- pedia lists are manually compiled by humans, but actually they are answers to semantic queries, and therefore perfectly suited for a system like ours.
In addition, the featured Wikipedia lists undergo a review process in the community, based on, besides other attributes, comprehensiveness.
For our ground truth, we automatically extracted the list of entities from the Wikipedia list pages.
This leaves us with 10 topics and a total of 2,367 corresponding entities in our ground truth\footnoteref{foot:repro-tool}.

For all of these tasks we manually generated queries in our query language corresponding to the semantics of the topics.
We relied on using the interactive query suggestions of our user interface, but did not fine-tune our queries towards the results.
An automatic translation from natural language to our query language is part of future work (see section \ref{sec:conclusions}).
We want to stress that our goal is not a direct comparison to systems that participated in the tasks above.
For that, input, collection and relevance judgments would have to be perfectly identical.
Instead, we want to show that our system allows to construct intuitive queries that provide high quality results for these tasks.

We first evaluated the impact of our context decomposition from Section 3.3 (\emph{contexts}) on result quality, by comparing it against two simple baselines: taking each sentence as one context (\emph{sentences}) and taking each section as one context (\emph{sections}).
%\begin{table}[ht]
%{\renewcommand{\baselinestretch}{1.3}\normalsize
%\hspace*{-1.5mm}
%\begin{tabular}{|l||c|c|c|c|c|}
%\hline
%          &    \FP &    \FN &   \prec  &   \rec  &    \f1   \\\hline
%sections  &  6.890 &     19 &   4.7\% &    81.6\%&    8.2\% \\\hline          
%sentences &    392 &     38 &   39.2\% &   65.0\%&    36.8\% \\\hline
%contexts  &    297 &     36 &   44.9\% &   66.5\%&   45.6\% \\\hline 
%\end{tabular}}
%\caption{Sum and averages over all TREC queries for Broccoli on sections, sentences and contexts.%
%\label{table:nonlp-nlp}}
%\end{table}
Table \ref{table:quality-results} shows that compared to sentences, our contexts decrease the (large) number of false-positives significantly for all benchmarks. %, resulting in an increased precision.
% Compared to the sentence level the number of false-negatives decreases as well.
% Due to a feature of the context decomposition that utilizes previously recognized lists, like those in HTML documents or on Wikipedia pages, even the number of false-negatives decreases: 
% List items are appended to the sentence preceding the list prior to SCI+SCR. 
For the TREC benchmark even the number of false-negatives decreases.
This is the case because our document parser pre-processes Wikipedia lists by appending each list item to the preceding sentence (before the SCI+SCR phase).
These are the only types of contexts that cross sentence boundaries and a rare exception.
For the Wikipedia list benchmark we verified that this technique did not cause any results that are in the lists from which we created the ground truth.
%The sentence level evaluation shows comparable recall but naturally lower precision.
Since the sentence level does not represent a true superset of our contexts we also evaluated on the section level. 
% Since using sentences also led to a lower recall in addition to the expected lower precision, we also evaluated on sections. 
We can observe a decrease in the number of false-negatives (a lot of them due to random co-occurrence of query words in a section) which does not outweigh the drastic increase of the number of false-positives. 
Overall, context decomposition results in a significantly increased precision and F-Measure, which confirms the positive impact on the user experience that we have observed.
%\footnote{F-Measure increase is highly statistically significant (\emph{p-value} of $0.002$ for two-sided Fisher's randomization test \cite{citeulike:fisher}) \TODO{also check this for new queries}.} 

Considering the ranking related measures in Table \ref{table:quality-results} we see a varying influence for the context based approach.
The number of cases where ranking quality improves, remains unchanged or decreases is roughly balanced.
This looks surprising, especially since the increase in F-measure is significant, but the reason is simple.
%The measures in Table \ref{table:quality-results} are highly dependent on a good \emph{ranking}.
So far our system uses simplistic ranks, determined by mere term frequency.
We plan to improve on that in the future; see Section \ref{sec:conclusions}.
We want to stress the following though.
Most semantic queries, including all from the TREC and SemSearch benchmark, have a small set of relevant results.
We believe that for such queries the quality of the result set as a whole is more important than the ranking within the result set.
Still, for the TREC benchmark, R-precision on contexts is $0.62$ and, for the SemSearch benchmark, mean average precision is $0.45$.
The best run from the TREC 2009 Entity Track when restricted to the English Wikipedia had an R-precision of $0.55$ as reported in \cite[Table 10]{DBLP:conf/cikm/BronBR10}.
The best result for the SemSearch List Search Track was a mean average precision of $0.279$ \cite{balog2011ntnu}.
Again, these results cannot be compared directly, but they do provide an indication of the quality and potential of our system.

\vspace{1mm}
\subsection{Error analysis}\label{sec:experiments:analysis}

To identify areas where our system can be improved we manually investigated the reasons for the false-positives and false-negatives when using contexts.
We used the TREC benchmark for this, because it has a reasonable number of queries and relevance judgments that still allow a costly manual inspection of the results.
We defined the following error categories.
For false-positives:
(FP1) a true hit which was \emph{missing} from the ground truth;
(FP2) the words in the context have a \emph{different meaning} than what was intended by the query;
(FP3) due to an error in the \emph{ontology};
(FP4) a mistake in the \emph{entity recognition};
(FP5) a mistake by the \emph{parser}.
(FP6) a mistake in our \emph{context decomposition}.
%(FP7) other, new category, don't know.
For false-negatives:
(FN1) there seems to be \emph{no evidence} for this entity in the Wikipedia based on the query we used. It is possible that the fact is present but expressed differently, e.g., by the use of synonyms of our query words;
(FN2) the query elements are \emph{spread} over two or more sentences;
(FN3) a mistake in the \emph{ontology};
(FN4) a mistake in the \emph{entity recognition};
(FN5) a mistake by the \emph{parser};
(FN6) a mistake in our \emph{context decomposition}.
%(FN7) other.

\def\m#1{\makebox[6.7mm]{#1}}
\begin{table}[ht]
\vspace{2mm}
{\renewcommand{\baselinestretch}{1.3}\normalsize
\hspace*{3mm}
\begin{tabular}{|c||c|c|c|c|c|c|c|}
\hline
\m{\#FP} & \m{FP1} & \m{FP2} & \m{FP3} & \m{FP4} & \m{FP5} & \m{FP6} \\\hline
297      &    55\% &    11\% &    5\%  &    12\% &    16\% &    1\%   \\\hline
\end{tabular}
\par
\vspace{2mm}
\hspace*{3mm}
\begin{tabular}{|c||c|c|c|c|c|c|c|}
\hline
\m{\#FN} & \m{FN1} & \m{FN2} & \m{FN3} & \m{FN4} & \m{FN5} & \m{FN6} \\\hline
36       &    22\% &    6\%  &  26\%   &    21\% &     16\% &    8\%  \\\hline
\end{tabular}}
\caption{Breakdown of errors by category.
% for the false-positives (top) and the false-negatives (bottom).
\label{table:error-categories}}
\end{table}

Table \ref{table:error-categories} provides the percentage of errors in each of these categories.
The high number in FP1 is great news for us:
many entities are missing from the ground truth but were found by Broccoli.
Errors in FN1 occur when full-text search with our queries on whole Wikipedia documents does not yield hits, independent from our contexts. 
Tuning queries or adding support for synonyms can decrease this number.
% Paragraph?
FP2 and FN2 comprise the most severe errors. 
They contain false-positives that still match all query parts in the same context but have a different meaning and false-negatives that are lost because contexts are confined to sentence boundaries. 
Fortunately, both numbers are quite small.

The errors in categories FP and FN 3-5 depend on implementation details and third-party components.
The high number in FN3 is due to errors in our current ontology, YAGO.
A closer inspection revealed that, although the facts in YAGO are reasonably accurate, it is vastly incomplete in many areas (e.g., the \emph{acted-in} relation contains only one actor for most movies).
Preliminary experiments suggest that switching to Freebase \cite{DBLP:conf/sigmod/BollackerEPST08} in the future will solve this and improve the results considerably (see section \ref{sec:conclusions}).
To mitigate the errors caused by entity recognition and anaphora resolution (FP4+FN4), a more sophisticated state-of-the-art approach is easily integrated.
Parse errors are harder. 
Assuming a perfect constituent parse for every single sentence, especially those with flawed grammar, is not realistic. 
Still, those errors do not expose limits of our approach.
We hope to enable SCI+SCR without a full-parse in the future (see Section \ref{sec:conclusions}). 
% Paragraph?
The low number of errors due to our context decomposition (FP6+FN6) demonstrates that our current approach (Section \ref{sec:preproc:nlp}) is already pretty good. 
Fine-tuning the way we decompose sentences might decrease this number even further.

%FN1 is due to lack of evidence in the collection and FN3 is due to ontology errors, two aspects which we cannot do much about here.
%FN1 includes those where the query didn't fit, e.g. words part should have used synonyms. 
%FP2: This is where contexts didn't help to avoid FPs. The words had a different meaning but still occurred in the same context, however, there was no error in the decomposition. Luckily a low number.
%FN2: These are the results which we lose, no matter how perfect our SCI+SCR (+ontology and entity recognition) are, so it is great news that this is so low.
%This suggests that context decomposition has the potential for drastically reducing the number of false-positives (see \TODO{rename +corrections?} stats with correction in Table \ref{table:nonlp-nlp}), without compromising recall much.\\
%FP/FN 3/4: We use YAGO out of the box and still very simple ER. So these are mostly easily correctable, at least to a large extent.\\
%FP/FN 5: Errors in the third party constituent parser, which caused wrong contexts. A different parser might give better results.\\
%FP/FN 6: Context-Decomposition only causes few FPs/FNs showing that it already works pretty well.
%Further improving our context decomposition is still a worthwhile direction for future research, especially to remedy mistakes made by the parser.

Naturally, an evaluation should not treat entities missing in the ground-truth in the same way as actual errors.
Table \ref{table:quality-results-corrected} provides quality measures for our benchmark based on sentences and contexts under three conditions:
(\emph{original}) evaluation based on the original TREC ground-truth;
(\emph{+missing}) with the entities from FP1 added to the ground truth;
(\emph{+correct}) with the errors leading to FP and FN 3,4,5 corrected.

\begin{table}[ht]
\vspace{2mm}
{%\renewcommand{\baselinestretch}{1.3}\normalsize
\hspace*{-1.5mm}
\begin{tabular}{|l|l||c|c|c|c|}
\hline
                              &            &     \f1  &   \pten  &  \rprec  &   \map   \\\hline\hline %&  \ndcgr  \\\hline\hline
\multirow{2}{*}{Sentences}    &   original &  $0.37$  &  $0.58$  &  $0.62$  &  $0.46$  \\             %&  $0.52$  \\
                              &   +missing &  $0.55$  &  $0.77$  &  $0.76$  &  $0.60$  \\\hline       %&  $0.66$  \\\hline
\multirow{3}{*}{Contexts}     &   original &  $0.46$  &  $0.58$  &  $0.62$  &  $0.46$  \\             %&  $0.55$  \\
                              &   +missing &  $0.65$  &  $0.79$  &  $0.77$  &  $0.62$  \\             %&  $0.70$  \\
                              &   +correct &  $0.86$  &  $0.94$  &  $0.92$  &  $0.85$  \\\hline       %&  $0.87$  \\\hline
\end{tabular}
\caption{Quality measures on TREC 2009 queries for three different levels of corrections.%
\label{table:quality-results-corrected}}}
\end{table}

The numbers for \emph{+correct} show the high potential of our system and motivate further work correcting the respective errors.
As argued in the discussion after Table \ref{table:error-categories}, many corrections are easily applied, while some of them remain hard to correct perfectly.

% When adding the missing entities to the ground truth, R-precision rises to $0.77$. 

%While the measures in Table \ref{table:quality-results} are purely dependent on a good \emph{ranking}, our system has not yet been optimized in that regard.
%Instead, ranks are determined by mere term frequency.
%On the one hand, ranking is not as important for many semantic queries, especially for those from the TREC benchmark. 
%The goal is to retrieve the whole set of relevant entities. 
%Comparing members of that set, one is rarely more relevant than the other.
%Instead, it is more important to decide which entities belong to the result and which do not. 
%Even a system that produced a perfect ranking on all entities had to decide where to cut-off the list in order to generate a proper answer.
%On the other hand, poor ranking motivates future work. 
%Especially for semantic queries with much larger result sets than those from the TREC benchmark (e.g. for queries resembling Wikipedia lists like the list of drug related deaths), it is convenient and sometimes necessary to see the best known entities on top and there is no reason why our approach should not be combined with a sophisticated ranking. 
\vspace{1mm}
\section{Conclusions and Future Work}\label{sec:conclusions}

We have presented Broccoli, a search engine for the interactive exploration of combined text and ontology data.
We have described the index, the natural language processing, and the user interface behind Broccoli.
% And we have provided evidence that Broccoli is indeed fast and easy to use, and that it gives search results of good quality.
And we have provided reproducible evidence that Broccoli is indeed fast and gives search results of good quality.

So far, we have implemented all the basic ideas we deemed necessary to provide a convincing proof of concept.
Based on this work, there are a lot of interesting directions for future research.

The underlying ontology plays a major role for our system.
By switching from YAGO to Freebase we expect a great improvement of the overall quality through a better coverage of relations and thus proposals and results (see Tables \ref{table:error-categories} and \ref{table:quality-results-corrected} in the previous section).
Our current approaches to entity recognition and anaphora resolution work well, but it might be possible to further improve result quality by incorporating more elaborate state-of-the-art approaches.
This would also allow the system to be more easily applied to other collections than Wikipedia (our current heuristics rely on its structure, see Section \ref{sec:preproc:er}).
Integrating simple inference heuristics could help to reduce the number of errors that are caused by facts that are spread over several sentences.
A high-quality sentence decomposition \emph{without} the need for an expensive and error-prone full parse should further increase result quality.
While query times are already low, optimized query processing and clever caching strategies have the potential to further improve speed.
%Usability of our system should be validated with a user study of our UI and the whole system.
To investigate how to best approach performance and quality improvements, an evaluation of Broccoli on a larger, web-like collection should provide valuable insights.
Automatically transforming natural language queries into our query language could help users that are accustomed to keyword queries in constructing their queries.
Finally, a user study of our UI and the whole system is an important next step.

%* Index: so far we have implemented our main idea (co-occurence lists for prefixes and entities). There is much potential for optimization: using compression techniques, special treatment for very long index lists; optimization of the filter procedure.
%
%* Caching: dito.
%
%* Entity recognition and coreference resolution: simple heuristics so far making use of particularities of Wikipedia. Switch to general system.
%
%* NLP: we consider out context decomposition a relatively shallow form of NLP. But so far, we compute it from an expensive full parse. We conjecture that computing it directly is both simpler and has the potential to give better results.
%
%* Inference: some mild form of intference is necessary for some queries, develop it.
%
%* User interface: \TODO{name one nice-to-have feature which we do not have yet}
%
%* Actually use for information extraction by interactively constructing (very) 	good queries for certain relations, and then running them to gather fact triples.
%
%* Support for values.
%
%* User study.

\vspace{2mm}
\section*{Acknowledgments}
\vspace{3mm}
This work is partially supported by the DFG priority program Algorithm Engineering (SPP 1307) and by the German National Library of Medicine (ZB MED).

%We want to thank Claudius Korzen for his valuable help in the laborious and time-consuming manual evaluation of our query results.

%\begin{figure*}[ht]
%\the\textwidth
%\hrule
%\vspace{1mm}
%\hrule height 5pt
%\vspace{1mm}
%\hrule
%\caption{A figure with a horizontal line}
%\end{figure*}


\bibliographystyle{abbrv}
\bibliography{broccoli}  % sigproc.bib is the name of the Bibliography in this case

\end{document}